\documentclass[aps,superscriptaddress,twocolumn]{revtex4}

\def\url#1{}
\usepackage{hyperref}
\usepackage{url}
\usepackage{graphicx,graphics,epsfig}   
\usepackage{dcolumn}    
\usepackage{bm}         
\usepackage{amsmath}    
\usepackage{verbatim}   
\usepackage{color}      
\usepackage{times}
\usepackage{natbib}
\usepackage{amsmath,amsfonts,amssymb,graphics,graphics,color,times}
\usepackage{bm}

\usepackage{latexsym}
\usepackage{amsmath}
\usepackage{amssymb}
\usepackage{amsfonts}
\usepackage{amsthm}
\usepackage{mathrsfs}
\usepackage{color,verbatim,graphics}
\DeclareMathAlphabet{\mathrsfs}{U}{rsfs}{m}{n}
\DeclareMathAlphabet{\mathpzc}{OT1}{pzc}{m}{it}
\DeclareMathAlphabet{\matheus}{U}{eus}{m}{n}
\DeclareMathAlphabet{\mathbbold}{U}{bbold}{m}{n}

\def\r1{\textbf{r}}

\newcommand{\ba}{\begin{eqnarray}}
\newcommand{\ea}{\end{eqnarray}}
\newcommand{\ban}{\begin{eqnarray*}}
\newcommand{\ean}{\end{eqnarray*}}
\newcommand{\be}{\begin{equation}}
\newcommand{\ee}{\end{equation}}

\newcommand{\ket}[1]{|#1\rangle}

\begin{document}

\title{A charged quantum dot micropillar system for deterministic light matter interactions}

\author{P. Androvitsaneas}
\affiliation{Centre for Quantum Photonics, H.H. Wills Physics Laboratory, University of Bristol,
Tyndall Avenue, Bristol, BS8 1TL, United Kingdom}

\author{A. B. Young}
\affiliation{Department of Electrical and Electronic Engineering, University of Bristol,
Merchant Venturers Building, Woodland Road, Bristol, BS8 1UB, UK}

\author{C. Schneider}
\affiliation{Technische Physik, Physikalisches Institut and Wilhelm Conrad R\"ontgen-Center for Complex Material Systems, Universit\"at W\"urzburg, Am Hubland, 97474 W\"urzburg, Germany}

\author{S. Maier }
\affiliation{Technische Physik, Physikalisches Institut and Wilhelm Conrad R\"ontgen-Center for Complex Material Systems, Universit\"at W\"urzburg, Am Hubland, 97474 W\"urzburg, Germany}

\author{M. Kamp}\affiliation{Technische Physik, Physikalisches Institut and Wilhelm Conrad R\"ontgen-Center for Complex Material Systems, Universit\"at W\"urzburg, Am Hubland, 97474 W\"urzburg, Germany}

\author{S. H\"ofling}\affiliation{Technische Physik, Physikalisches Institut and Wilhelm Conrad R\"ontgen-Center for Complex Material Systems, Universit\"at W\"urzburg, Am Hubland, 97474 W\"urzburg, Germany}
\affiliation{SUPA, School of Physics and Astronomy, University of St Andrews, St Andrews, KY16 9SS, United Kingdom}

\author{S. Knauer}
\affiliation{Centre for Quantum Photonics, H.H. Wills Physics Laboratory, University of Bristol,
Tyndall Avenue, Bristol, BS8 1TL, United Kingdom}

\affiliation{Department of Electrical and Electronic Engineering, University of Bristol,
Merchant Venturers Building, Woodland Road, Bristol, BS8 1UB, UK}\author{E. Harbord}
\affiliation{Centre for Quantum Photonics, H.H. Wills Physics Laboratory, University of Bristol,
Tyndall Avenue, Bristol, BS8 1TL, United Kingdom}

\author{C. Y. Hu}
\affiliation{Department of
Electrical and Electronic Engineering, University of Bristol,
Merchant Venturers Building, Woodland Road, Bristol, BS8 1UB, UK}

\author{J. G. Rarity}
\affiliation{Department of
Electrical and Electronic Engineering, University of Bristol,
Merchant Venturers Building, Woodland Road, Bristol, BS8 1UB, UK}

\author{R. Oulton}
\affiliation{Centre for Quantum Photonics, H.H. Wills Physics Laboratory, University of Bristol,
Tyndall Avenue, Bristol, BS8 1TL, United Kingdom}
\affiliation{Department of Electrical and Electronic Engineering, University of Bristol,
Merchant Venturers Building, Woodland Road, Bristol, BS8 1UB, UK}

\begin{abstract}
Quantum dots (QDs) are semiconductor nanostructures in which a three dimensional potential trap produces an electronic quantum confinement, thus mimicking the behaviour of single atomic dipole-like transitions. However unlike atoms, QDs can be incorporated into solid state photonic devices such as cavities or waveguides that enhance the light-matter interaction. A near unit efficiency light-matter interaction is essential for deterministic, scalable quantum information (QI) devices. In this limit, a single photon input into the device will undergo a large rotation of the polarization of the light field due to the strong interaction with the QD. In this paper we measure a macroscopic ($\sim6^o$) phase shift of light as a result of the interaction with a negatively charged QD coupled to a low quality-factor (Q$\sim290$) pillar microcavity. This unexpectedly large rotation angle demonstrates this simple low Q-factor design would enable near deterministic light-matter interactions.
\end{abstract}

\maketitle


The deterministic, lossless exchange of energy  between charged QDs and single photons has been shown as the potential building block for a full range of components required for QI and quantum communication ~\cite{PhysRevLett.92.127902, PhysRevB.78.085307,PhysRevA.78.032318}. A deterministic light-matter interaction would give both the ability to switch the photon state with a high fidelity, as well keeping photon loss near zero (high efficiency). To achieve these simultaneously, it is essential that all the photon energy that couples to and from the quantum emitter must do so almost exclusively within a well-defined electromagnetic mode, where one can input/collect single photons. Input/output coupling efficiency is parameterised by the $\beta$-factor, the ratio between the rate of coupling of the dipole to this well-defined mode compared to the total coupling rate of the dipole to all available electromagnetic modes, including leaky ones. 

Great success has been had in approaching this limit in several systems, including photonic crystal (PC) waveguides ~\cite{Lund-Hansen:2008zr} and photonic nano-wires ~\cite{Claudon:2010fk}. For optical cavities, however, this limit has proven difficult to approach. Light-matter interaction strengths in the "strong coupling" regime  have been achieved for high Q-factor pillar microcavities~\cite{nat-432-7014} and in photonic crystal cavities ~\cite{Yoshie:2004uq}, and could show high fidelity switching. However, the input/output mode is usually not well-defined in high Q-factor cavities: the escape rate to and from a well-defined input channel is similar to the escape rate to leaky cavity modes (CMs). These leaky modes arise either from the intrinsic design of the structure or from fabrication imperfections, putting a limit on the efficiency of high Q-factor microcavities where the escape rate into the input/output mode is slow by design. However, in a low Q-factor pillar the cavity lifetime is very short. Thus one may easily design a high $\beta$-factor structure with a well-defined input/output mode, a crucial advantage ~\cite{Barnes:2002kx}.

\begin{figure}
\centering
\includegraphics[width=0.48\textwidth]{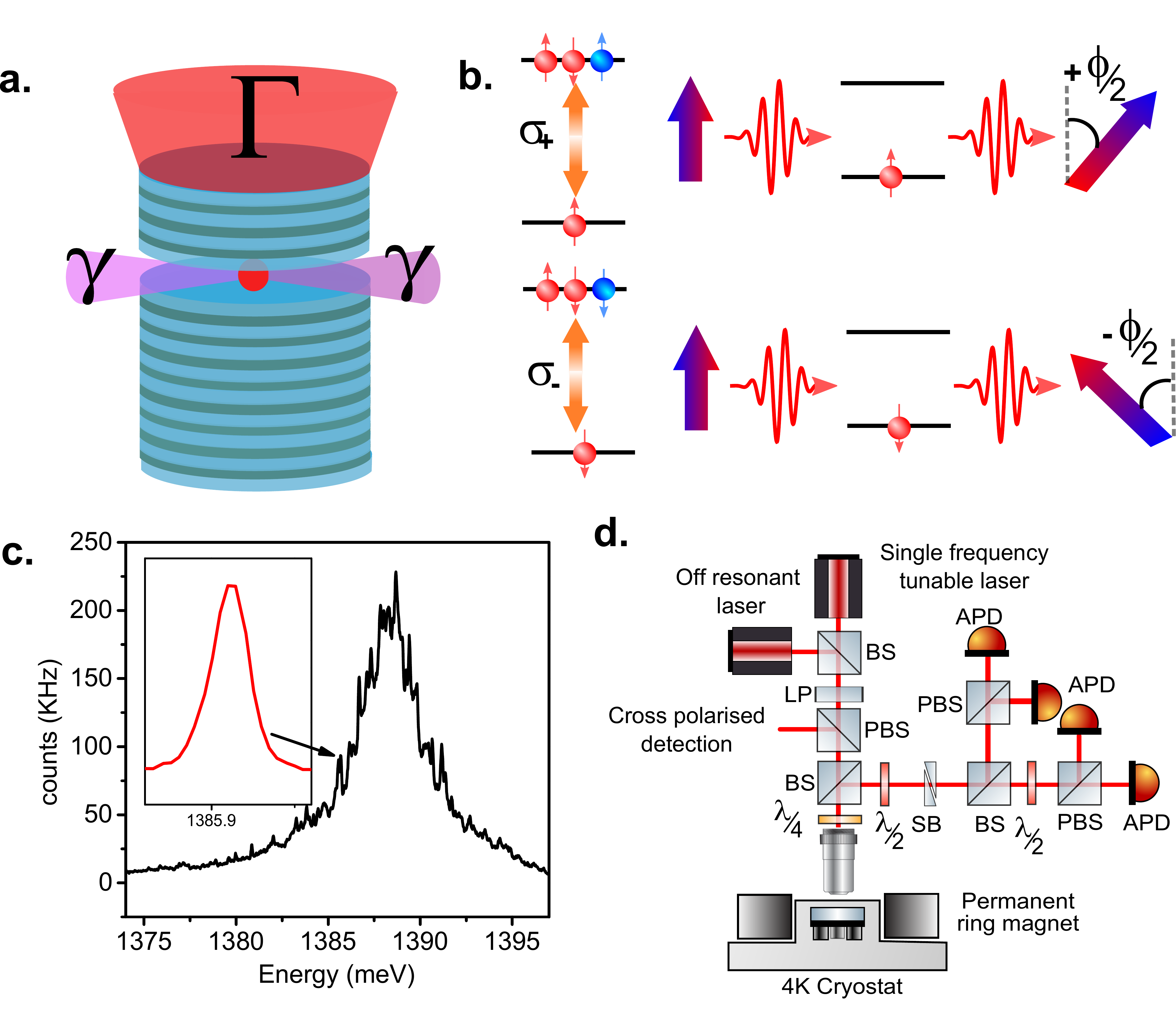}
\caption{ ({\bf a}) Schematic of the QD micropillar system with the available decay channels. ({\bf b}) Shows the available circular transitions for a negatively charged QD and the corresponding excitons created, along with the corresponding rotation of linearly polarised coherently scattered photons. ({\bf c}) Photo luminescence (PL) spectrum of the micropillar under consideration at T=12 K. The CM may be seen and the QD on top of the half maximum point of the CM. Inset: PL spectrum of the QD under lower off-resonant excitation power. ({\bf d}) Schematic of the experiment. The laser polarisation is set with a linear polariser (LP). The measured light is split on a non-polarising beamsplitter (BS) into two different measurement paths. The first path, for the identification of the transitions, is selected using a polarising beamsplitter (PBS) for the cross-polarised resonant scattering (detection path on the left). The second path is incident on a phase shift interferometer setup (on the right), where a quarter waveplate ($\lambda/4$, QWP) and a half waveplate ($\lambda/2$, HWP) are used to rotate the light to the correct measurement basis and a soleil-babinet compensator (SB) is used to account for any birefringence present and to calibrate our interferometer. Avalanche photodiodes (APDs) are used to record the signal.}\label{fig:1}
\end{figure}

The $\beta$-factor is directly linked to the competition between the rates of coherent and incoherent interaction present in these structures. The coherent coupling rate ($\Gamma$) is related to the parameter $g$, which represents the dipole cavity field coupling rate in the Jaynes-Cummings Hamiltonian ~\cite{Jaynes:1963kl}. In the case where the dipole is resonant with the cavity, this leads to a modified rate of emission given by $\Gamma=4g^2/\kappa$, where $\kappa$ represents the decay rate of the CM. The incoherent fraction is parameterised by $\gamma$ and $\gamma^*$, where $\gamma$ represents the rate at which the dipole radiatively couples to other available non-CMs (see Fig.\ref{fig:1}.a), and  $\gamma^*$ represents the pure dephasing rate of the dipole. The $\beta$ factor is now defined as $\beta=\frac{\Gamma}{\Gamma+\gamma+\gamma^*}$, the ratio of the rate of coherent interaction to the total interaction rate. In order to achieve high $\beta$-factors, one may either increase $g$ by decreasing the mode volume of the cavity (as $g\propto1/\sqrt{V_{eff}}$), or $\kappa$ which is inversely proportional to the Q-factor of the cavity. However one can also modify $\gamma$ geometrically, by reducing the number of available vacuum modes into which the dipole can decay ~\cite{Noda:2007uq}. This has been exploited in PCs to redistribute the rates of emission, enabling the design ~\cite{Lecamp:2007ve}, and realisation ~\cite{Lund-Hansen:2008zr} of high $\beta$-factor PC-waveguides. Here we show that this same redistribution occurs in low Q-factor micropillars. This contrasts with the conventional approach for micropillar cavities where the $\beta$-factor is increased via the strong enhancement of the decay rate into the CM (Purcell enhancement) ~\cite{Barnes:2002kx}.

A direct measure of the interaction strength is the magnitude of the phase shift induced on photons as they coherently scatter from the dipole transition~\cite{Young:2011uq}. It is a well-known quantum optics phenomenon ~\cite{Carmicheal_book} that in the limit where $\beta\sim1$ the light will experience a $\pi$-phase shift relative to the incident light ~\cite{Kochan:1994uq,Zumofen:2008fk}, the maximum possible in this configuration. In fact, it has been shown that as long as $\beta>0.5$ then the maximum possible phase shift of $\pi$ will always be observed ~\cite{Hofmann:2003fk}. Thus a high fidelity operation (i.e. a $\pi$ phase shift) will always be observed for $\beta >0.5$, while increasing the $\beta$-factor further increases the efficiency. Thus high efficiency and fidelity may be achieved when $\beta\sim1$.


In this work we consider a negatively charged QD containing a dopant electron. This has spin-selective transitions. If the excess electron is in the spin-up (-down) state, only $\sigma^+$ right ($\sigma^-$ left) circular polarized light can scatter from the dipole (see Fig.\ref{fig:1}.b). Accordingly, if the light is in a superposition of an interacting part and a non-interacting part, the induced phase shift will be picked up by the interacting component, while the non-interacting part has no phase shift. For example if vertically $\ket{V}$ polarised light scatters off a charged QD, the interacting part will acquire a phase shift $\phi$ dependent on the strength of the interaction, and the orientation of the QD spin  (i.e. $\ket{V}\ket{\downarrow}\rightarrow(e^{i\phi}\ket{\sigma^-}-\ket{\sigma^+})\ket{\downarrow}$, and $\ket{V}\ket{\uparrow}\rightarrow(\ket{\sigma^-}-e^{i\phi}\ket{\sigma^+})\ket{\uparrow}$) ~\cite{Atature:2007zr, J.Berezovsky12222006,Reiserer:2013ly}. This phase shift now maps onto a rotation along the linear polarization plane. This is known as the spin dependant Kerr (Faraday) rotation $\phi_r$, with $\phi_r=\phi/2$. Thus, for $\beta>0.5$ the maximum $\phi_r=\pi/2$ will be achieved, i.e. a rotation from $\ket{V}$ to $\ket{H}$~\cite{Hofmann:2003fk}. 

\begin{figure*}
\centering
\includegraphics[width=0.9\textwidth]{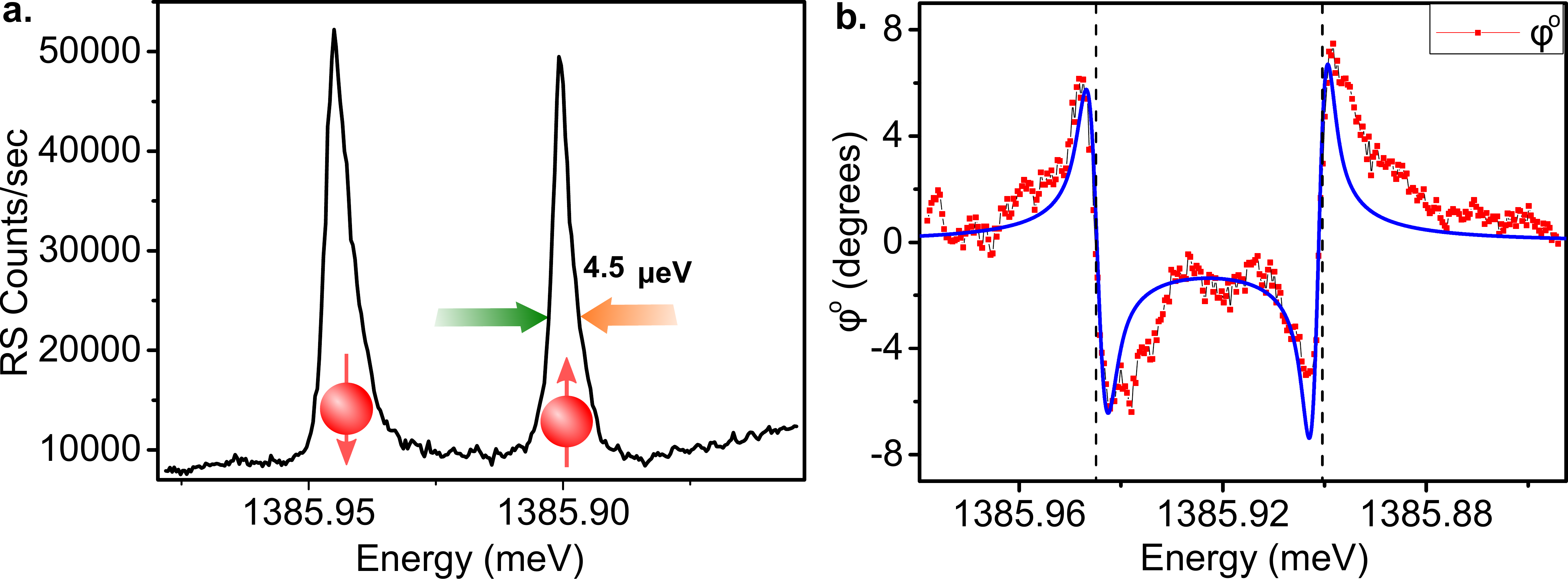}
\caption{({\bf a}) Photon counts using the resonant scattering technique detected in the cross-polarised arm. The two circular transitions are indicated, with arrows to indicate the spin orientation. ({\bf b}) The experimental data for the phase shift (red points) overlaid with the theoretical fit (blue line) after incorporating jitter and accounting for the thermalization between the spin states. The dashed lines correspond to the resonant frequency for each transition.}\label{fig:2}
\end{figure*}

We study a QD incorporated into the centre of a low Q-factor micropillar cavity, with CM Q-factor $\sim$ 290. The QD transition we examine in this work can be seen in Fig.\ref{fig:1}.c, where it is spectrally detuned by $\sim$2.7 meV with respect to the CM resonance (slightly below the half maximum of the CM). In order to define two circularly polarised transitions we introduce a magnetic field (500 mT) along the growth direction (Faraday geometry). The cross polarised resonant scattering (RS) signal of the Zeeman split doublet is shown in Fig.\ref{fig:2}.a, with an observed scattering response over $\sim$4.5 $\mu$eV. In order to measure the phase we perform polarisation analysis of the total RS signal, detecting simultaneously horizontal ($\ket{H}$), vertical ($\ket{V}$), diagonal ($\ket{D}$), and anti-diagonal ($\ket{A}$) (see Fig.\ref{fig:1}.d). The phase shift, $\phi $, can then be obtained using:

\be
sin\phi=\frac{D-A}{2\sqrt{HV}}
\ee

The measured phase shift response is shown in Fig.\ref{fig:2}.b., where a maximum phase shift of $\phi\sim6^o$ is observed, corresponding to a Kerr rotation of $\phi_r\sim3^o$. The measured Kerr rotation angle is similar in magnitude to that reported by Arnold {\it et al.} ~\cite{Arnold:2015bh}. The similarity in phase shift values is surprising since the Q-factor value of $\sim$2000 used in Ref. ~\cite{Arnold:2015bh} is an order of magnitude larger. Naively, the only difference between the high and low Q-factor micropillar would be the factor of 10 increase in photon loss rate ($\kappa$), resulting in a reduction of $\Gamma$ by an order of magnitude. Therefore, the measured phase shift from the low-Q micropillar would be of the order of mrad ~\cite{:/content/aip/journal/apl/88/19/10.1063/1.2202393}, assuming approximately the same value of $g$ in both cases. The macroscopic phase shift we have demonstrated indicates that this description of the system is inadequate.

\begin{figure}
\centering
\includegraphics[width=0.48\textwidth]{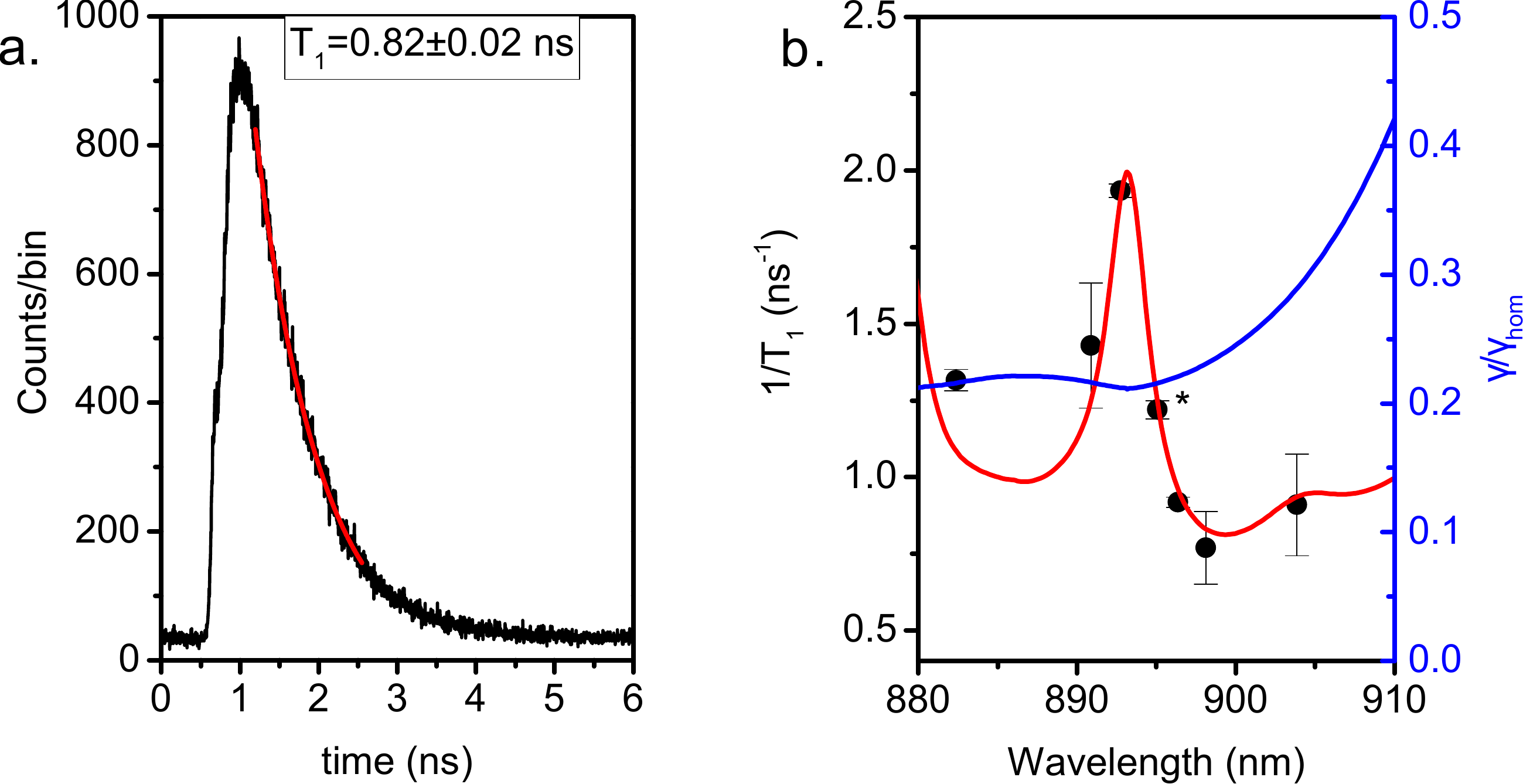}
\caption{ ({\bf a}) The measured lifetime of the QD transition under study. ({\bf b}) The black circles represent the $1/T_1$ quantity. The black star next to the black circle in the middle denotes the QD under study. The lifetimes where measured under pulsed p-shell excitation. $\Gamma_t/\gamma_{hom}$ was simulated and the simulated value for $1/T_1$ is plotted fitting for $T_{1hom}=710$ $ps$ (red curve).  $T_{1hom}=710$ $ps$ is in agreement with lifetime measurements performed on similar wavelength QDs without an etched structure from the same waffer. The blue curve is the ratio $\gamma/\gamma_{hom}$ as it is simulated for this structure.}\label{fig:3}
\end{figure}

In order to better understand this result we need to examine the sources of decoherence in our system. Fig.\ref{fig:3}.a. shows the QD lifetime measured under pulsed p-shell excitation showing a lifetime of 0.82$\pm$0.02 ns, corresponding to a Fourier transform-limited linewidth of $\Gamma_t=(\Gamma+\gamma)\sim0.8$ $\mu$eV. Clearly, this is significantly narrower than the 4.5 $\mu$eV measured using RS in Fig.\ref{fig:2}.a. The measured RS linewidth corresponds to the convolution of the Fourier limited spectral response, pure dephasing and any ``spectral jitter" present. Previous studies of QD RS have shown this broadening to be almost exclusively a product of spectral wander (jitter) as a result of charge and spin noise. By measuring the RS linewidth at a rate of 10's KHz almost transform-limited RS linewidths have been observed ~\cite{Kuhlmann:2015kx}. This suggests that the contribution from pure dephasing, occurring on time scales shorter than the QD radiative lifetime, is not significant. We confirm this in our case by measuring the first and second order correlation function (g$^{(1)}(t)$, g$^{(2)}(t)$ see Supplementary), of the cross-polarised RS photons. We increase the input laser intensity and measure a decay in the g$^{(1)}(t)$ corresponding to a coherence time of $\sim 5$ns, indicating $\gamma^*\ll\Gamma,\gamma$. Further, measurement of the autocorrelation ($g^{(2)}(t)$) reveals bunching of the signal on timescales of order $\mu$s, indicative of the spectral wander, in agreement with Kuhlmann et al. ~\cite{ Kuhlmann:2015kx}.


The effect of spectral wander will be to significantly wash out the observed phase features in Fig.\ref{fig:2}.b. A stochastic model may be used to describe this. We assume the QD transition has a $\Gamma_t=0.8$ $\mu$eV, with the resonance moving about a central frequency following a Gaussian profile, which agrees with the RS line-shape in Fig.\ref{fig:2}.a. By applying this model to the phase shift data in Fig.\ref{fig:2}.b, using the value of $\Gamma_t$ obtained from Fig.\ref{fig:3}.a., we can fit for one free parameter: $\Gamma$ (see supplementary information). The result is the fit (blue line) in Fig.\ref{fig:2}.b which gives $\Gamma\approx0.52$ $\mu$eV, resulting in a high $\beta$-factor of $\beta\approx0.65\pm0.03$. The only additional assumption here is that the spin is in a thermal state, and in the time averaged measurement we perform (1s integration time) each spin state is occupied only 50\% of the time. This effectively limits the maximum observable phase shift to $\phi=\pi/2$ for this measurement. This value for $\Gamma$ implies that the radiative decay rate into lossy modes is only $\gamma\approx0.28$ $\mu$eV. This is much smaller than the decay rate in homogeneous material (or free space) $\gamma_{hom}$ ($\sim1$ $\mu$eV), typically used as a value for $\gamma$ in these equations. This indicates a redistribution of the radiative decay channels from non-CMs into the CM, as has been previously measured in Ref.~\cite{PhysRevB.91.205310}, using high Q-factor pillar microcavities. 

This is perhaps not surprising when one considers the geometry of the micropillar cavity. In conventional atom-cavity QED where the cavities are macroscopic, the CM only subtends a small angle, leaving almost 4$\pi$ steradians of possible decay modes contributing to $\gamma$. This geometrical limitation leads to the assumption that $\gamma$ is similar to the free space decay rate $\gamma_{hom}$ ~\cite{Carmicheal_book}. Under this assumption, the only approach to achieve a high $\beta$-factor, is an enhancement of the decay rate  into the CM (Purcell enhancement) compared to homogeneous material  ~\cite{Gerard:1998cr, Barnes:2002kx, auffeves-garnier:053823}. However, for the wavelength-scale micropillar cavity used here, the geometry is very different to an atom-cavity.  A simple geometric approximation allows one to estimate that only $0.12\times4\pi$ steradians can escape below the critical angle from the side of the pillar (i.e. $\gamma=0.12\gamma_{hom}$). This is clearly an oversimplification, but regardless, it elucidates the underlying physics; in the real system our fits give a value $\gamma\sim0.3\gamma_{hom}$ (where $\gamma_{hom}=0.93$ $\mu$eV based on Fig.\ref{fig:3}.b). The simulated side losses (finite-difference time-domain (FDTD), blue curve in Fig.\ref{fig:3}.b), predict $\gamma\sim0.22\gamma_{hom}$. The discrepancy may be attributed to incoherent contributions from higher order modes or coupling through the bottom mirrors, but still the dominant contribution is from the side losses.

The fact that this micropillar shows a high $\beta$-factor for QDs close to resonance agrees well with previous calculations~\cite{Ho:2007dq}, and is confirmed in our simulations. Furthermore our simulations show that for a QD resonant with the CM, only $\sim$15$\%$ of the total emission from the QD radiatively couples out of the side of the micropillar ($\gamma=0.15\Gamma_{t}$).  The QD in question here is, however, slightly detuned from the cavity mode, $\sim$0.66 cavity linewidths. Naturally, the $\beta$ factor is frequency dependent: $\beta(\omega)=\frac{\Gamma(\omega)}{\Gamma(\omega)+\gamma(\omega)}$, where $\Gamma(\omega) = \frac{4g^2}{\kappa(1+(2(\omega_c-\omega)/\kappa)^2)}$, with $\omega_c$ and $\omega$ the cavity and emitter resonances respectively~\cite{Carmicheal_book}.  Hence one obtains a higher $\beta$-factor for an on-resonant QD ($\omega=\omega_c$): we predict for this micropillar $\beta(\omega_c)= 0.85$ (see Fig.\ref{fig:3}.b), compared to the studied QD with $\beta\approx 0.65$.  Note that the criterion for $\beta$-factor $>$0.5 holds even for a detuned QD, obtaining the maximum phase shift of $\pi$.

The frequency dependence of $\beta$ might imply, as the QD is strongly detuned, that the QD lifetime should increase dramatically. This has never been observed experimentally, as $\gamma$ is not independent of $\omega$. By strongly detuning a QD from the cavity mode, the DBRs now act as a stopband to create efficient waveguiding out of the side of the micropillars hence increasing $\gamma(\omega)$ relative to $\gamma(\omega_c)$, evidenced by the blue line in Fig.\ref{fig:3}.b (see also Supplementary).

This work demonstrates an ideal design for a micropillar: however, photonic design does not alleviate the problems with the properties of the emitter itself, such as phonon dephasing and spectral jitter.  Nevertheless, we have shown that pure dephasing of the transition due to phonons is minimal. Spectral wander is also not insurmountable. The time scales for the spectral jitter as a result of charge noise are around a 1-100 $\mu$s ~\cite{Kuhlmann:2015kx}. This is slow compared to typical spin coherence times of the order $\mu$s ~\cite{Greilich:2006uq}. Thus, single shot or time resolved Kerr rotation measurements ~\cite{:/content/aip/journal/apl/88/19/10.1063/1.2202393} as opposed to the time averaged experiment shown here, would allow observation of the maximum $\phi=\pi$ phase shift achievable with this particular QD-cavity combination. This should enable us to generate spin-photon entanglement within the spin coherence time with a ~50\% probability, compared to 0.003\% in the current state of the art ~\cite{Schaibley:2013fk}. Hence efficient spin-photon entanglement using such a low Q-factor design is indeed feasible.

In summary we show that a low Q-factor micropillar cavity meets the requirements for unity fidelity ($\beta$-factor of $> 0.5$), in an intrinsically high efficiency system. Strong light-matter interaction is inferred from a measurement of the input photon phase shift of $\sim 6^{o}$. Pure dephasing in this system is minimal, but the magnitude of the phase shift is nevertheless reduced due to easily quantifiable spectral jitter. By fitting to this spectral jitter and taking into account the thermal state of the spin and QD-cavity detuning we estimate the measured $\beta$-factor to be $\sim0.65$, with potential to allow a full $\pi$-phase shift of incident light if spectral jitter could be overcome.  No previous QD-microcavity design has been demonstrated that does not compromise efficiency for fidelity: for this simple to fabricate low Q-factor micropillar, fidelity and efficiency go hand-in-hand, thus enabling useful spin-photon entanglement devices.

\begin{acknowledgements}
The authors would like to thank H.F. Hofmann for helpful discussions. This work was funded by the Future Emerging Technologies (FET)-Open FP7-284743 [project Spin Photon Angular Momentum Transfer for Quantum Enabled Technologies (SPANGL4Q)] and the German Ministry of Education and research (BMBF) and Engineering and Physical Sciences Research Council (EPSRC) [project Solid State Quantum Networks (SSQN)].
\end{acknowledgements}

\appendix
\section{Methods}

We measure the resonant scattering (RS) from the QD using dark field microscopy techniques, where we tune a linearly polarised single frequency laser (1 second integration at each frequency) over the QD from the blue to the red side and collect the orthogonal linear polarisation at a power of $\sim1\%$ of the saturation power, we stress here that the RS signal has not been spectrally resolved. Initially the detection spectrum of the cross-polarised RS revealed large fluctuations of the intensity and the width of the scattering response. These are attributed to charge noise, where trapped charges undergo Coulomb interactions with the exciton transition, shifting its energy. Less than 10 pW of off-resonant laser (820 nm) was used to pacify the noise and reduce the observed broadening and fluctuations in intensity. This resulted in an observed RS linewidth of ~4.5 $\mu$eV. The QD has been identified as charged performing RS under different polarization bases. The Faraday geometry magnetic field was introduced by the use of a permanent ring magnet, which creates a homogeneous magnetic field perpendicular to the plane of the QD of 500 mT. The data in Fig.\ref{fig:2} were recorded simultaneously as shown in Fig.\ref{fig:1}.d. The lifetimes were measured under p-shell excitation with a micro photon devices (MPD) detector with $\sim$80 ps timing jitter and $\sim$4\% efficiency at $\sim$890 nm. The QDs measured in Fig.\ref{fig:3}.b. were chosen as they appeared bright, when the collected light was mode-filtered by a 5 $\mu$m core single mode fibre. This preferentially selects QDs aligned spatially in the centre of the micropillar, that are coupled into the fundamental CM with a well defined Gaussian spatial mode.

A commercial-grade simulator based on the FDTD method was used to perform the calculations~\cite{Lumerical}
  
\section{Device design and fabrication}
The system under consideration is a micropillar photonic structure. The sample has been fabricated via molecular beam epitaxy (MBE) and contains a $\lambda$-thick cavity surrounded by two distributed Bragg reflectors (DBRs). The DBR structure consists of 18 (5) bottom (top) AlAs/GaAs mirror pairs. A modulation doped low density In(Ga)As QD layer ($1.8$x$10^9$ cm$^{-2}$) has been grown in the middle of the cavity. The QDs have been grown spectrally close to the CM resonance at ~1388 meV. The asymmetric low-Q design ensures that the cavity field decays predominantly through the top mirror ($>90\%$). In this way a high QD emission extraction efficiency is expected for these systems. From a selection of different etched micropillar diameters one of a nominal diameter  $\sim2$ $\mu$m was selected. We find that the fundamental mode of the cavity appears at $1388$ meV  cavity and has a full width half maximum of $\sim4.1$meV corresponding to a Q-factor of $Q\sim290$. The fundamental mode is well separated from any higher modes that appear typically $\gtrsim16$meV to the blue side of the fundamental mode. The cumulative rate of losses into higher order modes are calculated to be not more than 3\% of the rate of emission into the fundamental mode. The high light collection efficiency is evident from measured QD count rates of 1 MCount/sec at saturation under resonant excitation.  
\bibliography{bib13desk}

\end{document}